\documentclass[conference]{IEEEtran}
\usepackage{cite}
\usepackage{amsmath,amssymb,amsfonts}
\usepackage{algorithmic}
\usepackage{amsmath, amssymb}
\usepackage[linesnumbered,ruled,vlined]{algorithm2e}
\usepackage{graphicx}
\usepackage{textcomp}
\usepackage{paralist}
\usepackage{tcolorbox}
\usepackage{enumitem}
\usepackage{xcolor}
\usepackage{dblfloatfix}
\usepackage{float}
\def\BibTeX{{\rm B\kern-.05em{\sc i\kern-.025em b}\kern-.08em
    T\kern-.1667em\lower.7ex\hbox{E}\kern-.125emX}}
\usepackage{acronym}
\usepackage{subcaption}
\usepackage{soul}

\acrodef{AI}[AI]{Artificial Intelligence}
\acrodef{ML}[ML]{Machine Learning}
\acrodef{5G CN}[5G CN]{5G Core Network}
\acrodef{5G-NR}[5G-NR]{5G New Radio}
\acrodef{5G}[5G]{Fifth Generation}
\acrodef{3GPP}[3GPP]{3rd Generation Partnership Project}
\acrodef{URLLC}[URLLC]{Ultra Reliable Low Latency Communications}
\acrodef{eMBB}[eMBB]{Enhanced Mobile Broadband}
\acrodef{mMTC}[mMTC]{Massive Machine-Type Communication}
\acrodef{O-RAN}[O-RAN]{Open Radio Access Network}
\acrodef{PCA}[PCA]{Principal Component Analysis}
\acrodef{IoT}[IoT]{Internet of Things}
\acrodef{RAN}[RAN]{Radio Access Network}
\acrodef{TIP}[TIP]{Telecom Infra Project}
\acrodef{RIC}[RIC]{RAN Intelligent Controller}
\acrodef{near-RT}[near-RT]{near-real-time}
\acrodef{non-RT}[non-RT]{non-real-time}
\acrodef{CU}[CU]{Central Unit}
\acrodef{DU}[DU]{Distributed Unit}
\acrodef{RU}[RU]{Radio Unit}
\acrodef{COTS}[COTS]{Commercial-Off-The-Shelf}
\acrodef{QoS}[QoS]{Quality of Service}
\acrodef{IaaS}[Iaas]{Infrastructure as a Service}
\acrodef{SLA}[SLA]{Service Level Agreement}
\acrodef{PaaS}[PaaS]{Platform as a Service}
\acrodef{SaaS}[SaaS]{Software as a Service}
\acrodef{RRC}[RRC]{Radio Resource Control}
\acrodef{KPI}[KPI]{Key Performance Indicator}
\acrodef{UE}[UE]{User Equipment}
\acrodef{SMO}[SMO]{Service Management and Orchestration}
\acrodef{LSTM}[LSTM]{Long Short-Term Memory}
\acrodef{O-CU}[O-CU]{Open Centralized Unit}
\acrodef{O-DU}[O-DU]{Open Distributed Unit}
\acrodef{MCS}[MCS]{Modulation and Coding Scheme}
\acrodef{SINR}[SINR]{Signal to Interference plus Noise Ratio}
\acrodef{CQI}[CQI]{Channel quality Indicator}
\acrodef{UL}[UL]{Uplink}
\acrodef{DL}[DL]{Downlink}
\acrodef{RF}[RF]{Random Forest}
\acrodef{PDU}[PDU]{Protocol Data Unit}
\acrodef{UE}[UE]{User Equipment}

\begin{document}

\title{FALCON: A Framework for Fault Prediction in Open RAN Using Multi-Level Telemetry}

\author{ 
\IEEEauthorblockN{Yaswanth Kumar LS$^\star$, Somya Jain$^\star$, Bheemarjuna Reddy Tamma$^\star$, and Koteswararao Kondepu$^\dag$}
\IEEEauthorblockA{
$^\star$ Indian Institute of Technology Hyderabad, India\\
$^\dag$ Indian Institute of Technology Dharwad, India\\
}
\IEEEauthorblockA{E-mail: \{cs24resch11007, cs23mtech12011, tbr\}@iith.ac.in, k.kondepu@iitdh.ac.in}
}

\maketitle
\begin{abstract}
\ac{O-RAN} has brought in deployment flexibility and intelligent RAN control for mobile operators through its disaggregated and modular architecture using open interfaces. 
However, this disaggregation introduces complexities in system integration and network management, as components are often sourced from different vendors. 
In addition, the operators who are relying on open source and virtualized components --- which are deployed on commodity hardware --- require additional resilient solutions as O-RAN deployments suffer from the risk of failures at multiple levels including infrastructure, platform, and RAN levels.
To address these challenges, this paper proposes FALCON, a fault prediction framework for O-RAN, which leverages infrastructure-,  platform-, and RAN-level telemetry to predict faults in virtualized O-RAN deployments. By aggregating and analyzing metrics from various components at different levels using AI/ML models, the FALCON framework enables proactive fault management, providing operators with actionable insights to implement timely preventive measures. The FALCON framework, using a Random Forest classifier, outperforms two other classifiers on the predicted telemetry, achieving an average accuracy and F1-score of more than $98\%$.

\end{abstract}


\section{Introduction}

As mobile networks become increasingly complex and dynamic, the need for intelligent and adaptive network management has become crucial. 
Traditional \ac{RAN} solutions, designed primarily for earlier generations of cellular networks, struggle to meet the 5G use cases requirements due to their closed and monolithic nature. Vendor lock-in and proprietary hardware hinder flexibility, making it difficult for mobile operators to introduce new features, integrate diverse technologies or respond to rapid changes in traffic patterns and service demands~\cite{oran2018}. 
%Review#1 Comments#1
Moreover, traditional RAN management relies on static, pre-defined configurations, lacking real-time optimization capabilities. 
As traffic demands fluctuate and dynamic services are introduced, the traditional RAN architectures lead to inefficiencies and increased operational costs. 
Fault management in traditional \ac{RAN} is reactive, and addresses only after they cause significant disruptions, which makes it challenging to ensure continuous service availability~\cite{Sihem2019}.

The new RAN architectures, namely Open RAN (\ac{O-RAN}) architecture offers a paradigm shift in how network services are deployed and managed~\cite{oran2018}. 
The O-RAN provides a significant advantage by disaggregating RAN functions, by introducing \ac{RIC} and open interfaces among network functions that foster collaboration and interoperability among multiple vendors while allowing for greater flexibility in network deployments using \ac{COTS} hardware and cloud native technologies.
\emph{Openness} in O-RAN facilitates innovation, enabling seamless integration of third-party apps and empowering the operators to leverage advanced \ac{AI}/\ac{ML} algorithms for intelligent RAN control and automation.

A typical \ac{O-RAN} 5G deployment requires diverse multi-vendor components, including \ac{COTS} hardware, virtualization technologies, open-source and proprietary software for O-RAN functions such as Near-RT RIC, \ac{O-CU}, \ac{O-DU}, third-party xApps/rApps, and \ac{SMO}. 
This diversity poses integration and management challenges. 
Unlike previous generations tightly integrated components, \ac{O-RAN} relies on multi-vendor solutions. 
While offering cost savings and flexibility, the disaggregated approach lacks built-in fault tolerance as these networks could be susceptible to faults, bugs, and misconfigurations and suffer from resource contentions and performance issues as shown in Table~\ref{table:1}. The link failures often require traffic rerouting, straining resources and reducing system stability. Ensuring resilience in this heterogeneous ecosystem requires advanced fault management, predictive analytics, and robust orchestration to mitigate challenges for widespread \ac{O-RAN} adoption.

\begin{table}[htb!]
\caption{Anomalous behaviors, their causes, and potential implications in virtualized O-RAN deployments}
\label{table:1}
\centering
\scalebox{0.84}{
\renewcommand{\arraystretch}{2}  % Increases row height (vertical spacing)
\begin{tabular}{|p{3cm}|p{3cm}|p{3cm}|}  % Set column width to 4cm for each column
\hline
\textbf{Anomalous behavior} & \textbf{Potential Faults/Issues} & \textbf{Potential Failures} \\
\hline
Sudden spike in server temperature, decrease in CPU clock rate & Inadequate cooling, thermal throttling due to high workload, malfunctioning of fans or cooling units & Performance degradation of \ac{O-RAN} NFs, CPU throttling leading to latency spikes, system shutdown, impact on UEs \\
\hline
Sudden increase in memory usage & Excessive logging by xApps or \ac{O-RAN} NFs, memory leaks in software components, misconfigured processes or memory allocation & Crash of \ac{O-RAN} NFs or containers, Disruption in UE sessions or service drops \\
\hline
High CPU usage, bit-rate spikes, increased traffic & Misconfiguration of routing policies, \ac{RAN} overload & \ac{UE} disconnectivity, crash of \ac{O-RAN} NFs\\
\hline
Reduced traffic, timeouts, \ac{PDU} re-establishments & Link faults on O-RAN interfaces, network congestion & Reduced throughput, Disruption in UE sessions or service drops \\
\hline
High CPU load, saturation of VSwitch port & VNF resource contention, high traffic volume, inefficient load balancing across NF instances & Application service crash, end-to-end service failures \\
\hline
\end{tabular}
}
\end{table}
                                                
To address such challenges in O-RAN deployments, a precise monitoring system is crucial. 
Such a system shall provide transparent telemetry data by ensuring trust and accountability without vendor bias when attributing faults. 
This approach empowers operators to take timely preventive measures, creating an end-to-end resilient network across infrastructure, platform, and application (RAN) levels. In this work, we collect multi-level telemetry in virtualized \ac{O-RAN} deployments, including infrastructure-level metrics (host-level telemetry), platform-level metrics (container-level telemetry), and application-level metrics (RAN telemetry).

This paper proposes a novel framework titled \textit{FALCON} which aims to achieve self-resilience in O-RAN systems using this multi-level telemetry. 
The \textit{FALCON} framework continuously collects and analyzes telemetry data at all levels, forecasting Key Performance Indicators (KPIs) to predict faults in advance. This proactive monitoring enables operators to implement preemptive measures, addressing issues before they escalate into larger disruptions, thus maintaining network stability and performance.

The \textit{FALCON} framework employs dimensionality reduction techniques combined with forecasting models to handle high-dimensional O-RAN telemetry data and thereby limits the focus only to the key features. It provides accurate and efficient anomaly predictions by classifying anomalous behavior to determine the type of fault, significantly reducing the resource overhead caused by the high-dimensional data of O-RAN telemetry.
This approach enhances scalability, enabling FALCON to efficiently process large volumes of high-dimensional telemetry data from complex O-RAN deployments while maintaining performance as networks grow.

The rest of the paper is as organized follows: Section~\ref{section:related_work} presents the related work, followed by the proposed FALCON framework in Section~\ref{proposed_work_section}. Section~\ref{experimental_setup_section} describes the experimental setup, while Section~\ref{performance_evaluation_section} presents the performance results. Finally, Section~\ref{conclusions_futurework_section} provides the conclusions and future directions.

\section{Related Work}
\label{section:related_work}
This section reviews existing works on resilience in cellular networks and cloud environments, and presents research gaps.

The authors of~\cite{Sihem2019} provided a comprehensive survey on fault management techniques used in traditional network deployments and also studied the impact of virtualization in network fault management. The authors of~\cite{Sauvanaud2016} used ML models to detect Service Level Agreement (SLA) violations in NFV environments and preliminary symptoms of SLAs violations so as to help the operators to locate anomalous VMs that caused SLA violations for applying appropriate recovery techniques in a proactive manner. In~\cite{mukh2023}, the authors presented different stages of fault management process by using classical ML and neural network-based deep learning. 

Live migration and recovery mechanisms are proposed in the literature to address failure of RAN and mobile core components. The authors of~\cite{Shunmu2022,ramanathan2024enabling} proposed live-migration schemes for both containerized mobile core and CU NFs in 5G. Atlas~\cite{xing2023enabling} ensures \ac{DU} resilience using proactive and reactive migrations through existing cellular mechanisms like handovers and cell reselection. Strategies such as pre-copy, post-copy, and hybrid migrations \cite{bhattacharyya2023towards} are used for improving resiliency of specific RAN component(s) in these works. , yet challenges persist in achieving end-to-end resiliency across infrastructure, platform, and RAN levels in virtualized O-RAN deployments. 

Misconfigurations are a significant source of faults in Open RAN systems. The authors of~\cite{YUNG2024} categorize misconfigurations into three main categories: integration and operation related, SDN/NFV related, and AI/ML related. This work emphasizes the role of AI/ML based detection techniques, including anomaly detection for KPI analysis and correlation analysis to identify conflicting xApps in O-RAN. Similarly, systems like SpotLight \cite{sun2024spotlight} uses advanced anomaly detection techniques such as \emph{JVGAN} and \emph{MRPI}, achieving high accuracy and efficient bandwidth usage. SpotLight further incorporates explainability tools like \emph{KFilter} and \emph{CausalNex} for root cause analysis. While effective for anomaly detection and localization, these solutions do not offer broader application (RAN) level and infrastructure level fault management solutions. Also these works focus mainly on detecting configuration errors or faults based on RAN or platform level KPIs collected, but not on forecasting potential faults and failures for taking proactive measures to make the system self-resilient. 

In summary, existing works in the literature excel in specific areas like cloud resource management, DU/CU recovery, and O-RAN anomaly detection, but they do not offer holistic solutions for predictive fault management across infrastructure, platform, and RAN levels. To address this gap, we propose FALCON, a holistic framework for virtualized O-RAN deployments. 

\section{Proposed FALCON Framework}
\label{proposed_work_section}

\begin{figure}[htb!]
    \centering
    \includegraphics[width=0.8\linewidth]{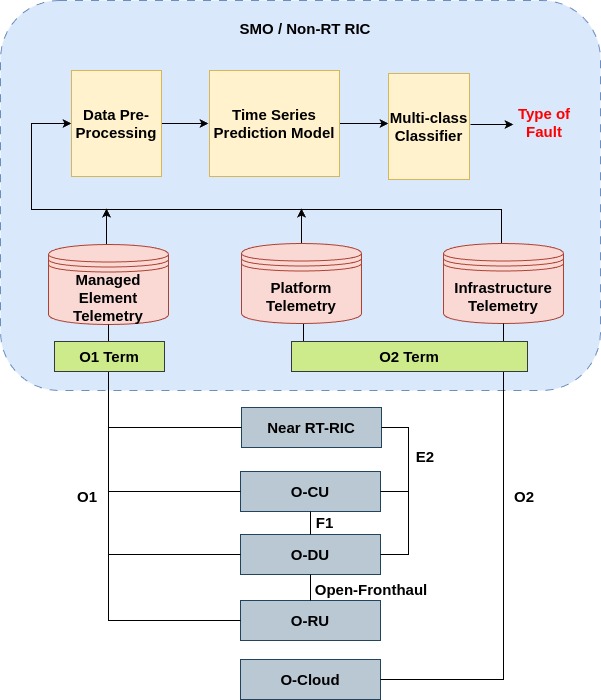}
    \caption{Solution Architecture of FALCON Framework.}
    \label{fig:FALCON}
\end{figure}

Fig.~\ref{fig:FALCON} shows the architecture of the proposed \textit{FALCON} framework. The O1 interface facilitates the operation and management (e.g., fault management) of O-RAN components, while the O2 interface manages the O-Cloud infrastructure and the life cycle of O-RAN network functions.
In the proposed framework, the \ac{SMO} periodically collects three types of telemetry data from the O-RAN system. It collects the managed element telemetry --- also referred as application-level or RAN telemetry --- from O-RAN components, including O-RU, O-DU, and O-CU, via the O1 interface. The platform telemetry and infrastructure telemetry are collected from the O-Cloud over O2 interface using exporters. The gathered telemetry data are stored in separate time-series databases within the SMO. 

%%%%%%%%%%%%%%%%%%%%%%%%%%%%%%%%%%%%%%%%%%%%%%%%%%%%%%%%%%%%
%%%%%%%%%%%%%%%%%%%%%%%%%%%%%%%%%%%%%%%%%%%%%%%%%%%%%%%%%%%%

In this work, we collect diverse telemetry data from \ac{O-RAN} components and the underlying O-Cloud to estimate network state.
For example, consider a scenario where \ac{RAN} telemetry shows increased packet loss (throughput degradation) and latency spikes. Analyzing only these metrics might lead to an incorrect diagnosis, such as partial link failures and network congestion faults. However, the network performance degradation may also happen due to host CPU frequency drops triggered by CPU temperature spikes. Hence, the FALCON framework considers the following multi-level telemetry data to prevent misdiagnosis. 

\begin{itemize}
\item RAN Telemetry: It includes alarms and events for fault management, current configuration settings of the managed elements, KPIs related to network health and user-specific metrics, power metrics, etc. 
\item Platform Telemetry: This includes container-level CPU metrics, memory usage, network statistics, filesystem metrics, disk I/O metrics, and more which help in identifying container-specific performance issues such as bottlenecks, misconfigurations, or resource limitations, enabling efficient resource allocation and performance optimization within the containerized environment.
\item Infrastructure Telemetry: This includes host-level CPU metrics, memory usage, network statistics, filesystem metrics, disk I/O metrics, and system health-related metrics which help to monitor the overall health of the host system by detecting hardware issues, excessive resource consumption, and performance degradation, ensuring the stability and reliability of the underlying O-Cloud infrastructure. 
\end{itemize}

The collected telemetry data is pre-processed, where data from different levels is integrated and imputed to handle missing values. FALCON employs dimensionality reduction techniques to handle potentially large dimensionality of telemetry data while preserving its essential features and allowing scalability. In addition, dimensionality reduction helps to reduce computational cost while predicting telemetry values. Since \ac{RAN} components are interlinked, the collected telemetry has both spatial and temporal dependencies. In order to preserve the spatial dependencies, telemetry's reduced features are given to forecaster model as a single input.

%%%%%%%%%%%%%%%%%%%%%%%%%%%%%%%%%%%%%%%%%%%%%%%%%%%%%%%%%%%%%%%%%
\begin{figure}[htb!]
    \centering
    \includegraphics[width=1.0\linewidth]{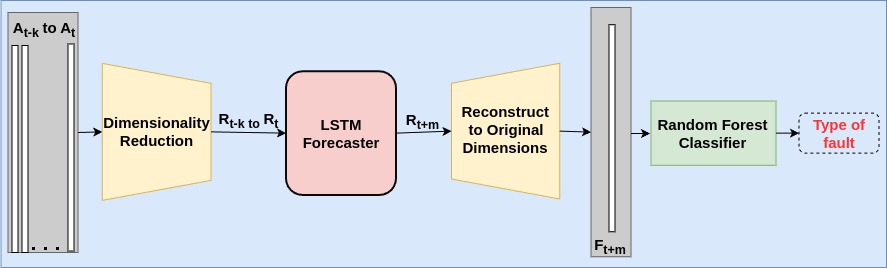}
    \caption{Detailed ML pipeline used in FALCON.}
    \label{fig:ML_architecture}
   % \vspace{-1em}
\end{figure}
%%%%%%%%%%%%%%%%%%%%%%%%%%%%%%%%%%%%%%%%%%%%%%%%%%%%%%%%%%%%%%%%%%%

From Fig.~\ref{fig:ML_architecture}, $\mathbf{A}_{t-k}$ to $\mathbf{A}_t$ denote the preprocessed telemetry data vector spanning from time step $(t-k)^{\text{th}}$ to $t^{\text{th}}$, where $k > m$ and $m$ is the future prediction step. \ac{PCA}, a dimensionality reduction technique, is applied to this vector to obtain a set of reduced features, $\mathbf{R}_{t-k}$ to $\mathbf{R}_t$. 
These reduced features are provided as an input to the \ac{LSTM} time-series forecasting model. The \ac{LSTM} Forecaster model processes the sequential input through layers to capture the temporal dependencies and relationships within the data. Specifically, the \ac{LSTM} model utilizes: (i) an \ac{LSTM} layer with input size of $\mid \mathbf{R}_{t} \mid$, $h$ hidden units, and some additional layers to learn the temporal and spatial patterns and (ii) a fully connected linear layer with $h$ input features and $\mid \mathbf{R}_{t} \mid$ output features.  

The \ac{LSTM} model forecasts the future values of the reduced features at $(t+m)^{\text{th}}$ time step. A single vector $\mathbf{F}_{t+m}$ is reconstructed to original dimensions from $\mathbf{R}_{t+m}$, and subsequently fed to a multi-class \ac{RF} classifier. Since, \ac{RAN}'s anomalous behavior depends not only on CPU/Memory usage but also on the traffic generated by the \ac{UE}s as well as number of \ac{UE}s connected to the network, \ac{RF} classifier is used as it makes decisions related to the type of fault based on lower/upper thresholds of multiple features.

\section{Experimental Setup}
\label{experimental_setup_section}
This section describes the experimental setup used to evaluate the performance of the proposed FALCON framework. It also includes details of the hardware and software tools used, how various faults are injected, what telemetry data is collected during the experiments, and the ML-based fault prediction pipeline.

%%%%%%%%%%%%%%%%%%%%%%%%%%%%%%%%%%%%%%%%%%%%%%%%%%%%%%%%%%%%%%%%
\begin{figure}[htbp]
        \centering
         \vspace{-1em}
        \includegraphics[width=1\linewidth]{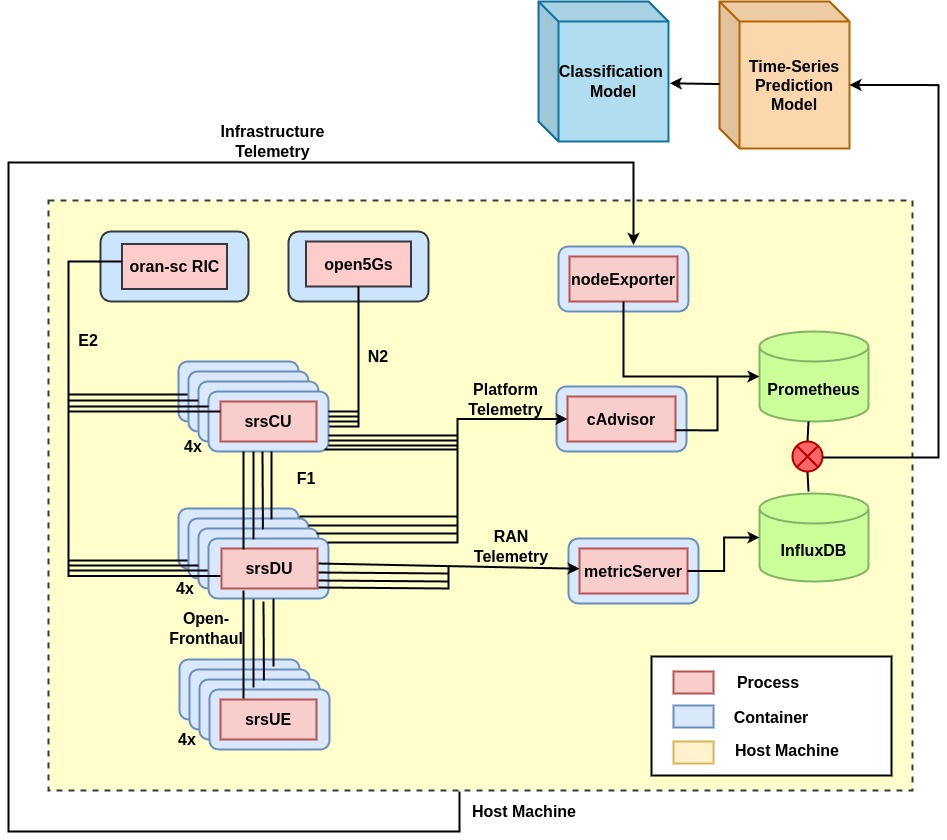}
        \caption{Experimental Testbed.}
        \label{fig:infocomExperimental_setup}
        \vspace{-1em}
    \end{figure}
%%%%%%%%%%%%%%%%%%%%%%%%%%%%%%%%%%%%%%%%%%%%%%%%%%%%%%%%%%%%%%%

\subsection{Testbed Setup}
We utilized a server equipped with two Intel Xeon ES-2690 v4 processors, providing a total of 54 cores, along with 64 GB of RAM and running Ubuntu 22.04 for our experimental setup. In this server, we integrate the O-RAN SC Near-RT RIC~\cite{oran_sc_RIC} together with the O-RAN components: srsCU, srsDU, and srsUE~\cite{srsRANProject}. 
The srsCU resources restricted with 3 CPU cores and 2 GB of RAM, while the srsDU had 3 CPU cores and 3 GB of RAM while other components are not resource restricted. Furthermore, Open5GS~\cite{open5GS} was deployed as the core network. Each component is deployed as a \emph{Docker container} to ensure modularity and scalability. In total, the testbed consists of four CUs, four DUs, and four UEs, each with a one-to-one mapping, as shown in Fig.~\ref{fig:infocomExperimental_setup} to create scenarios where only a subset of \ac{RAN} components are induced with stress to study its impact on other \ac{RAN} components.

Table~\ref{tab:telemetry-collection} shows the details of telemetry collection from the containers by using node exporter~\cite{node_exporter_hub}, cAdvisor ~\cite{cadvisor_prometheus}, and Metrics Server\cite{srsRANProject}. Prometheus~\cite{prometheus} stores collected data from \emph{cAdvisor} and \emph{Node Exporter} at a frequency of one second. 
In contrast, InfluxDB~\cite{influxDB} stores RAN telemetry from the Metrics Server, which receives updates from all \ac{DU}s every 100 milli-sec as shown in Fig.~\ref{fig:infocomExperimental_setup}. 

%%%%%%%%%%%%%%%%%%%%%%%%%%%%%%%%%%%%%%%%%%%%%%%%%%%%%%%%%%%%%%%%%
\begin{table}[htbp]
\caption{Details of Telemetry Collection in FALCON Framework}
\label{tab:telemetry-collection}
\centering
\renewcommand{\arraystretch}{1.5} 
\scalebox{0.80}{ 
\begin{tabular}{|p{1.8cm}|p{1.8cm}|p{5.5cm}|} 
\hline
\textbf{Level} & \textbf{Tools Used} &  \textbf{Type of Data Collected} \\ \hline
\textbf{RAN (DU) --- Managed Element Telemetry} & \textbf{Metric Server} collects metrics from DU &
\begin{minipage}[t]{5.5cm}
\begin{itemize}
    \item Number of active UEs
    \item Current total downlink bitrate
    \item Maximum total downlink bitrate
    \item Downlink bitrate
    \item Uplink bitrate
    \item Uplink \ac{MCS}
    \item Downlink \ac{MCS}
    \item Uplink \ac{SINR}
    \item \ac{CQI}
\end{itemize}
\vspace{0.1em}
\end{minipage} \\ \hline

\textbf{Platform (All Containers)} & \textbf{cAdvisor Exporter} collects metrics from DUs, CUs & 
\begin{minipage}[t]{5.5cm}
\begin{itemize}
    \item CPU usage
    \item Memory usage 
    \item Network statistics 
    \item Filesystem utilization
    \item Container metrics as per \cite{cadvisor_prometheus}
\end{itemize}
\end{minipage} \\ \hline

\textbf{Infrastructure (Host Machine)} & \textbf{Node Exporter} collects from the host system & 
\begin{minipage}[t]{5.5cm}
\begin{itemize}
    \item CPU usage 
    \item Memory utilization 
    \item Disk I/O metrics
    \item Filesystem statistics
    \item Network metrics such as packet \\transmission rates, errors
    \item Node temperature and power
    \item Additional telemetry as per \cite{nodeexporter}
\end{itemize}
\vspace{0.1em}
\end{minipage}  \\ \hline
\end{tabular}
}
\end{table}

\begin{figure}[htb!]
    \centering
    %\vspace{-2em}
    \includegraphics[width=0.95\linewidth]{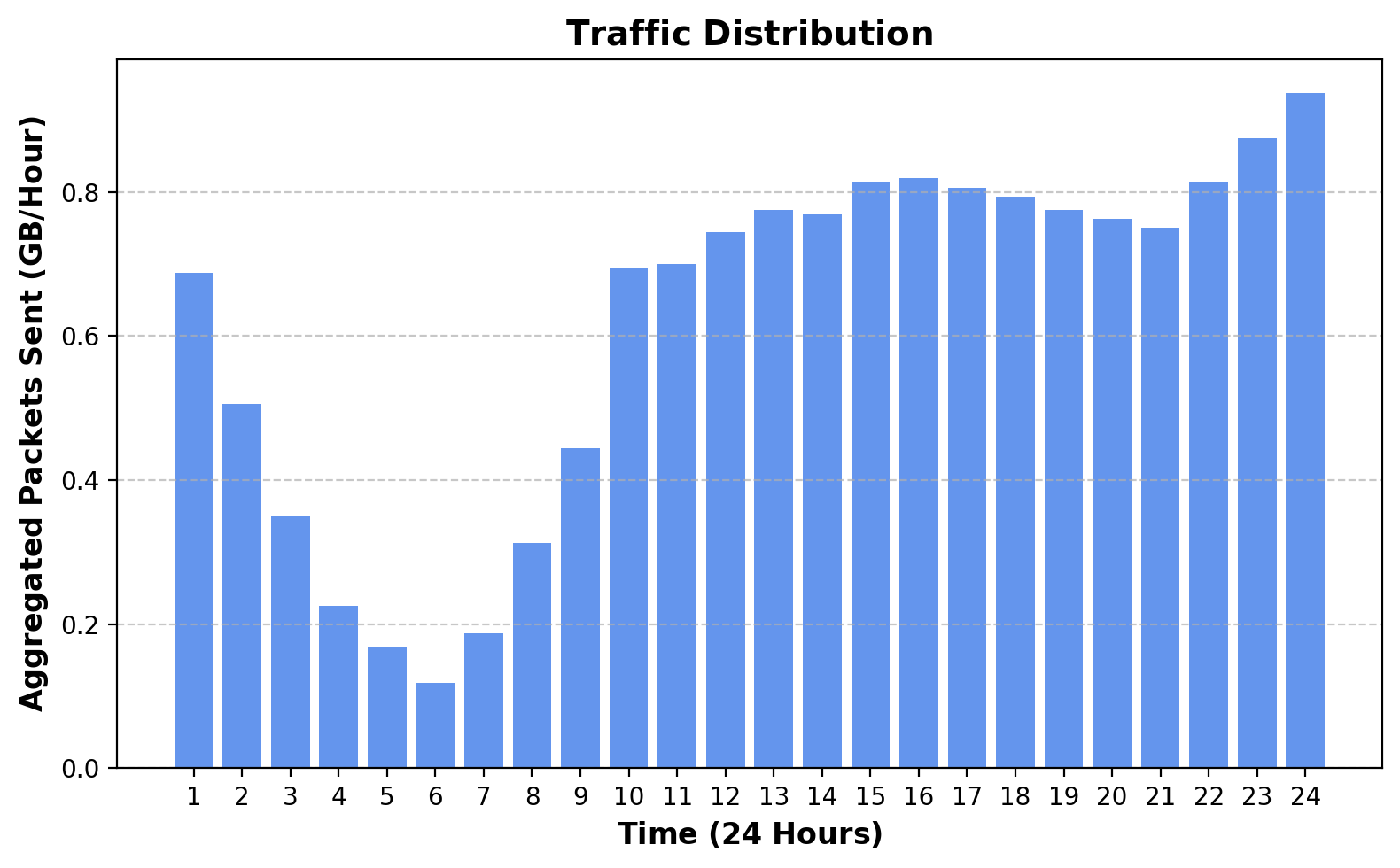}
    \caption{Traffic distribution generated by ping messages.}
    \label{fig:traffic_distribution}
    \vspace{-1em}
\end{figure}

\noindent\emph{Traffic generation:} We generate user traffic by sending ping messages every 100 milli-sec to the core network, with the packet size randomly varying for every 5 seconds in such a way that the aggregate user traffic per hour follows the distribution as shown in Fig.~\ref{fig:traffic_distribution} for both uplink and downlink. iPerf traffic is generated in the uplink direction from the users to fully occupy the channel with user traffic.

\noindent\emph{\ac{LSTM} Forecaster Configuration:} We configured the \ac{LSTM} model with \( \mid\mathbf{R}_t\mid = 10 \), \( h = 32 \) which were defined in Section~\ref{proposed_work_section} to create an \ac{LSTM} forecaster model with input size of $10$, $32$ hidden units, and two additional layers. Additionally a fully connected layer with $32$ input features and $10$ output features is attached to the two additional layers.

%%%%%%%%%%%%%%%%%%%%%%%%%%%%%%%%%%%%%%%%%%%%%%%%%%%%%%%%%%%%%%%%%%%%
%%%%%%%%%%%%%%%%%%%%%%%%%%%%%%%%%%%%%%%%%%%%%%%%%%%%%%%%%%%%%%%%%%%%
    
\subsection{Fault Injection Approach}

The fault injection process involves emulating different real-world faults in virtualized O-RAN deployments (refer to Table~\ref{table:1}) and collecting telemetry data at various levels for capturing anomalous system behavior due to these faults. 
In this work, we use tools like \textit{stress-ng}~\cite{stress_ng_123} to stress-test various instances of O-RAN components deployed in the considered testbed by forcing high CPU contentions and exhausting memory capacity. 
During this process, some packet loss may occur. Thus, to emulate the packet loss close to the real-time scenario, the \textit{tc} tool is used to introduce some packet losses on the interface connecting RAN instances in our setup (refer to Fig.~\ref{fig:infocomExperimental_setup}).  
%\clearpage
%\thispagestyle{empty}
\begin{algorithm}[htbp]
%\small
\caption{Fault Injection Procedure} % Use an unnumbered caption
\begin{minipage}[b]{0.93\columnwidth} 
\label{algo:fault}
\begin{algorithmic}[1]
% \REQUIRE $\mathcal{CU}=\{CU_0, CU_1, CU_2, CU_3\}$ $\mathcal{DU}=\{DU_0, DU_1, DU_2, DU_3\}$ \tcp*{Set of CUs and DUs}
\REQUIRE \( X = \mathcal{CU} \cup \mathcal{DU} \) \tcp*{Set of CUs and DUs to be monitored in the O-RAN deployment}

% \STATE \( X=\mathcal{CU} \cup \mathcal{DU} \) \tcp*{Set of all containers in the testbed}
\STATE Sample fault duration \( T_1 \sim \text{Exp}(\lambda) \mid T_1 \in [30, 90] \, \text{minutes} \) \tcp*{ $\text{Exp}(\lambda)$ is an exponential distribution}
\STATE Select fault type \( F \) using a multinomial distribution:  
    \[
     P_F(0)=0.3, \, P_F(1)=0.5, \, P_F(2)=0.1, \, P_F(3)=0.1
    \]
    \tcp*{\( F=0 \): Normal, \( F=1 \): CPU Stress, \( F=2 \): Memory Stress, \( F=3 \): Packet Loss}
\FOR{each container \( C \in X \)}
    \STATE Decide to inject stress using a Bernoulli distribution:
    \[
    P(\text{Stress} = 1) = 0.4,  P(\text{Stress} = 0) = 0.6
    \]
    \IF{\(\text{Stress} = 1\)}
        \IF{\( F = 1 \) (CPU stress)}
            \STATE Sample \textit{Start stress} \( \sim \mathcal{U}(0.4, 0.9) \) \tcp*{CPU stress is selected between 40\% and 90\%}
            \STATE Sample \textit{End stress} \( \sim \mathcal{U}(\text{Start stress}, 1.0) \) 
            \tcp*{End stress is greater than or equal to Start stress}
            \STATE Execute CPU stress in container \( C \) from the selected Start stress to End Stress for $T_1$ duration
        \ELSIF{\( F = 2 \) (Memory stress)}
            \STATE Sample \textit{Start stress} \( \sim \mathcal{U}(0.25, 0.35) \) \tcp*{Memory stress is selected between 25\% and 35\%}
            \STATE Sample \textit{End stress} \( \sim \mathcal{U}(\text{Start stress}, 0.6) \)
            \STATE Execute Memory stress in container \( C \) from the selected Start stress to End Stress for $T_1$ duration
        \ELSIF{\( F = 3 \) (Packet loss)}
            \STATE Sample \textit{Start stress} \( \sim \mathcal{U}(0.01, 0.03) \) \tcp*{Packet loss is selected between 1\% and 3\%}
            \STATE Sample \textit{End stress} \( \sim \mathcal{U}(\text{Start stress}, 0.05) \)
            \STATE Execute Packet loss in container \( C \) from the selected Start stress to End Stress for $T_1$ duration
        \ENDIF
    \ELSE
        \STATE No stress is injected into container \( C \) for \( T_1 \) duration
    \ENDIF
\ENDFOR
\STATE Repeat from Step 2 for the desired number of iterations

\end{algorithmic}
\end{minipage}
%\clearpage
\end{algorithm}

The various types of faults we inject are described in the following:

\begin{itemize}
\item \textbf{CPU stress test:} In this test, we stress the CPU cores assigned to different O-RAN components like CUs and DUs, and monitor the impact by collecting telemetry data at different levels as shown in Table~\ref{tab:telemetry-collection}. Depending on the CPU resource demands of various components, each of these components is expected to show some negative impact of insufficient CPU resources in terms of anomalous behavior at certain stress levels, which could be captured through the telemetry data being collected at various levels in the system.
     
\item \textbf{Memory stress test:} In this test, we stress the memory allocated to different CUs and DUs to mimic memory leaks and memory contentions. The negative impact is monitored by collecting telemetry data at different levels as shown in Table~\ref{tab:telemetry-collection}.
    
\item \textbf{Packet loss test:} In this test, we emulate partial link faults and congestion scenarios by introducing packet losses using the \textit{tc} tool.
\end{itemize}

To inject these faults systematically in the considered setup, we follow the procedure given in Algorithm \ref{algo:fault}, where the selected O-RAN components are induced with stress for certain time windows.

Note that the data is collected continuously with normal and faults injection scenarios, with $65,765$ samples used for training and testing. As shown in Fig.~\ref{fig:ML_architecture}, we created an ML pipeline with $k=60$ and $m=5$, where $k$ represents the considered backward steps and $m$ represents prediction step. The retraining of the considered ML pipeline depends on the topology changes, addition or deletion of features, etc.

\subsection{Classifier Selection}
We also evaluated two more classifiers (XGBoost and AdaBoost) along with \ac{RF} for predicting faults based on the forecasted  telemetry in the FALCON framework. 

\section{Performance Evaluation}
\label{performance_evaluation_section}
The proposed \emph{FALCON} framework was evaluated using stratified \textit{5}-fold cross-validation to predict future telemetry and classify it as normal or fault types (CPU Stress, Memory Stress, and Packet Loss). The accuracy of \emph{FALCON} shown in Table~\ref{table:apm} is obtained using $10$ PCA features from $403$ initial features, which leads to $\approx94.2\%$ reduction in features and the average RMSE value of $0.05789$ is achieved for predicted features after inverse  \ac{PCA} (before de-normalization).

%%%%%%%%%%%%%%%%%%%%%%%%%%%%%%%%%%%%%%%%%%%%%%%%%%%%%%
\begin{table}[htb!]
\centering
\caption{Averaged Performance of FALCON Across All Folds}
\label{table:apm}
\begin{tabular}{|l|c|}
\hline
\textbf{Metric} & \textbf{Value} \\
\hline
\textbf{Accuracy} & 98.73\% \\
\hline
\textbf{F1-Score} & 98.71\% \\
\hline
\textbf{Macro Average Precision} & 96.70\% \\
\hline
\textbf{Macro Average Recall} & 97.00\% \\
\hline
\textbf{Macro Average F1-Score} & 96.56\% \\
\hline
\textbf{Weighted Average Precision} & 98.81\% \\
\hline
\textbf{Weighted Average Recall} & 98.73\% \\
\hline
\textbf{Weighted Average F1-Score} & 98.71\% \\
\hline
\end{tabular}
\end{table}

\subsection{Interpretation of Results}
\label{interpretation_results_section}
FALCON demonstrated exceptional performance across all folds, achieving an average accuracy of $98.73\%$. Table~\ref{table:confusion} presents the average confusion matrix, which shows that the model accurately classified most of the samples for each class with minimal misclassifications. Notably, the \emph{normal} and \emph{CPU stress} classes were classified with very good accuracy, while a few misclassifications were observed for the \emph{memory stress} and \emph{packet loss} classes.

%%%%%%%%%%%%%%%%%%%%%%%%%%%%%%%%%%%%%%%%%%%%%%%%%%%%%%%%%%%%%%%%%%%%
\begin{table}[htb!]
\centering
\caption{Average Confusion Matrix Across All Folds}
\label{table:confusion}
\scalebox{0.90}{ % Adjust this value if needed to scale down the table
\begin{tabular}{|c|c|c|c|c|}
\hline
\textbf{True/Pred} & \textbf{Normal} & \textbf{CPU Stress} & \textbf{Memory Stress} & \textbf{Packet Loss} \\
\hline
\textbf{Normal} & 4693.2 & 2.0 & 7.8 & 1.2 \\
\hline
\textbf{CPU Stress} & 45.6 & 7012.4 & 50.4 & 7.4 \\
\hline
\textbf{Memory Stress} & 1.0 & 6.4 & 859.4 & 2.0 \\
\hline
\textbf{Packet Loss} & 8.0 & 3.6 & 31.2 & 421.4 \\
\hline
\end{tabular}
}
\end{table}

\begin{figure}[htb!]
    \centering
   % \vspace{-1em}
    \includegraphics[width=1.0\linewidth]{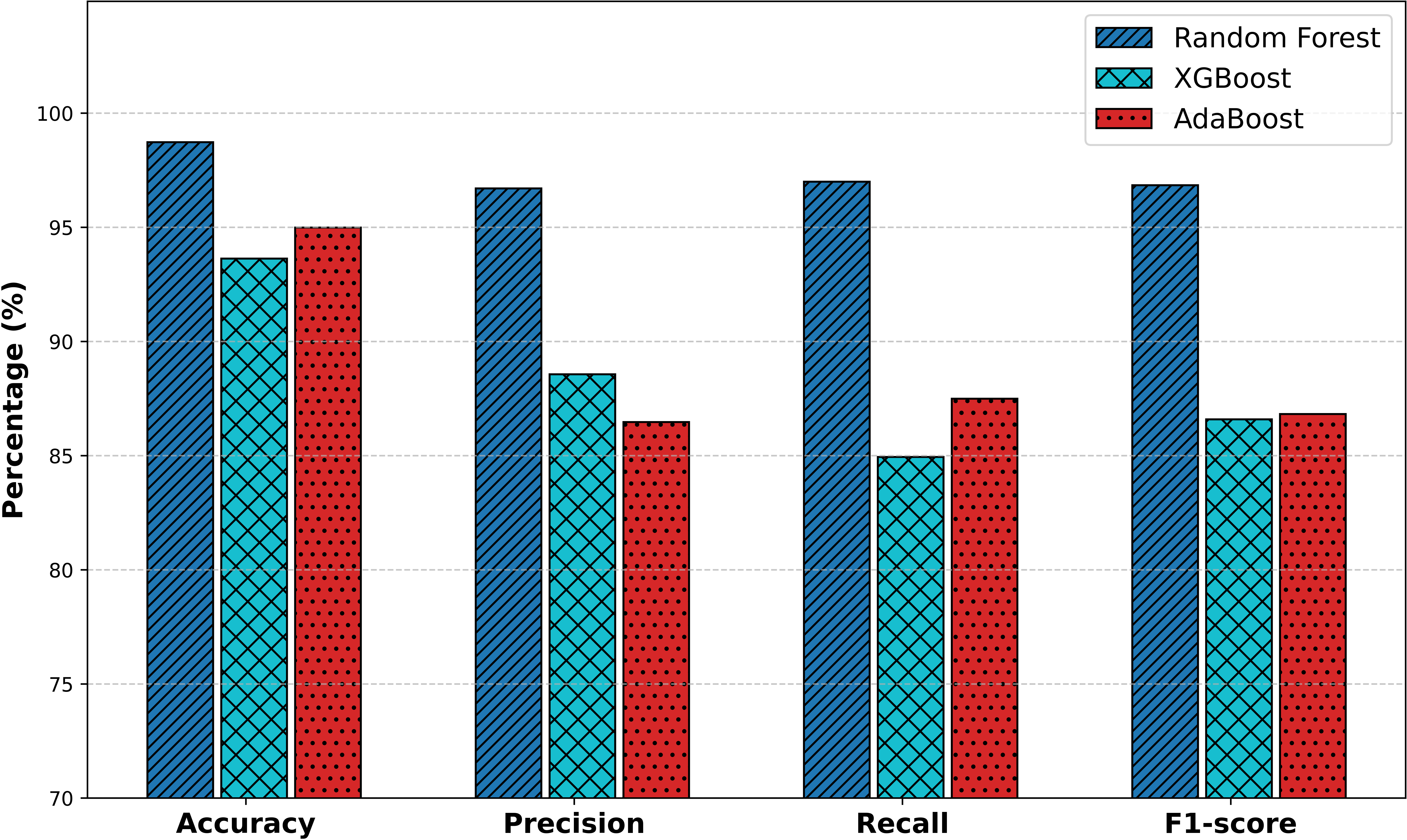}
    \caption{Average Performance of Different Classifiers in FALCON Framework Across All Folds.}
    %\vspace{-1em}
    \label{fig:performance}
\end{figure}

Fig.~\ref{fig:performance} shows \emph{Accuracy}, \emph{Precision}, \emph{Recall}, and \emph{F1-score} of FALCON for three different classifiers. Random Forest, which is employed by FALCON by default, outperforms XGBoost and AdaBoost due to its ensemble nature, which allows it to generalize well to unseen data. XGBoost and AdaBoost struggled with outliers/noisy data and got skewed towards those anomalies as both focus on learning from mistakes made by previous iterations which leads to overfitting issues.

%%%%%%%%%%%%%%%%%%%%%%%%%%%%%%%%%%%%%%%%%%%%%%%%%%%%%%%%%%%%%%
%%%%%%%%%%%%%%%%%%%%%%%%%%%%%%%%%%%%%%%%%%%%%%%%%%%%%%%%%%%%%%%

Table~\ref{table:report} provides the average classification report, highlighting FALCON's outstanding precision, recall, and F1-scores across all classes. The model achieved very good scores for the \emph{normal} and \emph{CPU stress} classes, with slightly lower but still good scores for \emph{memory stress} and \emph{packet loss} classes. Specifically, the F1-score for the CPU stress class was $99.18\%$, reflecting the framework's reliability even for challenging classifications.

%%%%%%%%%%%%%%%%%%%%%%%%%%%%%%%%%%%%%%%%%%%%%%%%%%%%%%%%%%%%%%%%%%%%%%
\begin{table}[htb!]
\caption{Average Classification Report Across All Folds in FALCON Framework}
\label{table:report}
\centering
\setlength{\arrayrulewidth}{0.3mm} % Increase table line width
\setlength{\tabcolsep}{5pt} % Increase column spacing
\begin{tabular}{|c|c|c|c|c|}
\hline
\textbf{Class} & \textbf{Precision} & \textbf{Recall} & \textbf{F1-Score}\\
\hline
0 (Normal) & 0.9888 & 0.9976 & 0.9931\\
1 (CPU Stress) & 0.9982 & 0.9854 & 0.9918\\
2 (Memory Stress) & 0.9109 & 0.9891 & 0.9471\\
3 (Packet Loss) & 0.9702 & 0.9077 & 0.9304\\
\hline
\end{tabular}
\end{table}

%\section{Results}

\section{Conclusions and Future Work}
\label{conclusions_futurework_section}
Designing a fault-proof system that uses anomalies to predict faults in advance and takes corrective measures is challenging due to the high dimensionality of telemetry data in virtualized O-RAN deployments. 
FALCON, our proposed framework, addresses this challenge by using \ac{PCA} for dimensionality reduction and \ac{LSTM} forecaster for KPI forecasting. It successfully forecasts anomalies to predict faults up to 5 seconds (or 5 time steps) in advance, enabling proactive decision-making by the operators. FALCON demonstrated strong performance with an average accuracy of $98.73\%$ and F1-score of $98.71\%$, across all folds of the stratified n-fold cross-validation. The confusion matrix and classification report show minimal misclassifications, highlighting its robustness.

Future work involves creation of a dataset on an emulated testbed consisting of a large number of O-RAN components and UEs to mimic real-world deployment scenarios with a greater number of faults and simultaneous fault cases. 
Subsequently, we plan to focus on root cause analysis to localize faults to be able to take preventive measures in a timely manner.

\section*{Acknowledgments}
This work was partially supported by Intel India and Visvesvaraya PhD Scheme, Meity, Govt. of India. The authors would like to thank Abdulla Ovais and Michael Suguna Kumar V for their help in designing \ac{ML} pipeline of the FALCON framework.

%\bibliographystyle{IEEEtran}
%\bibliography{biblio}

\end{document}